%% file: PurificationPaper.tex
\documentclass[5p,preprint,sort&compress,times,twocolumn]{elsarticle}

\usepackage{graphicx}

\usepackage{subfigure}

\usepackage{multicol}

\usepackage{longtable}
\usepackage{threeparttable}

\usepackage{amssymb}

\newcommand{\antinu}{anti-neutrino}
\newcommand{\ls}{LS}
\newcommand{\argon}{$^{39}$Ar} %
\newcommand{\krypton}{$^{85}$Kr}
\newcommand{\radon}{$^{222}$Rn}
\newcommand{\nitrogen}{N$_2$}

\newcommand{\BeNeutrino}{$^{7}$Be}

\newcommand{\ThPb}{$^{212}$Pb}
\newcommand{\UPb}{$^{214}$Pb}
\newcommand{\lead}{$^{210}$Pb}
\newcommand{\Lead}{lead}
\newcommand{\Radon}{radon}
\newcommand{\Pb}{lead}
\newcommand{\Rn}{radon}
\newcommand{\degC}{$^{\circ}{\rm C}$}

\newcommand{\CPb}{lead-carbon}

\newtheorem{Hypothesis}{Hypothesis} 

\journal{Nuclear Instruments and Methods in Physics Research, A}

\begin{document}


\begin{frontmatter}

\title{Laboratory Studies on the Removal of Radon-Born Lead from KamLAND's Organic Liquid Scintillator}
\input{AuthorListNIM}





\begin{abstract}

The removal of radioactivity from liquid scintillator has been studied in preparation of a low background phase of KamLAND.  This paper describes the methods and techniques developed to measure and efficiently extract \Radon\ decay products from liquid scintillator.   
We report the radio-isotope reduction factors obtained when applying various extraction methods. During this study, distillation was identified as the most efficient method for removing \Radon\ daughters from liquid scintillator.  
\end{abstract}

\end{frontmatter}


\input{Intro}

\input{Procedure}

\input{Results}

\input{Conclusion}


\vskip 0.2in
\noindent
{\bf Acknowledgments}
\vskip 0.1in
\noindent
This work was supported in part by the U.S. Department of Energy (DOE) grant DE-FG02-01ER41166,
the Japanese Ministry of Education, Culture, Sports, Science and Technology,
the World Premier International Research Center Initiative (WPI Initiative), MEXT, Japan,
DOE contract number DE-AC02-05CH11231, and other DOE grants from individual KamLAND institutions.  The authors would like to thank Dr. Li Cang of Selecto Scientific Inc., and 
Xiong Xin Dai and Nick Jelly (Oxford University) of the SNO collaboration.

\bibliographystyle{model1-num-names}
\bibliography{master.bib}

\end{document}

%% file: AuthorListNIM.tex
%
%
\author[UA,atllnlnow]{G.~Keefer\corref{cor1}}
\cortext[cor1]{Corresponding author at: Lawrence Livermore National Laboratory, Livermore, California 94550, USA Tel.: +1 925 424 5094; Fax: +1 925 424 5512}
\ead{gregkeefer@llnl.gov}
\author[UA,atucdnow]{C.~Grant}
\author[UA,IPMU]{A.~Piepke}
\author[RCNS]{T.~Ebihara}
\author[RCNS]{H.~Ikeda} 
\author[RCNS,IPMU,aticrrnow]{Y.~Kishimoto}
\author[RCNS,atokayamanow]{Y.~Kibe}
\author[RCNS]{Y.~Koseki}
\author[RCNS]{M.~Ogawa} 
\author[RCNS]{J.~Shirai}
\author[RCNS]{S.~Takeuchi}
\author[CALTECH,atlanlnow]{C.~Mauger}
\author[CALTECH,atbnlnow]{C.~Zhang}
\author[UTChem]{G.~Schweitzer}
%
%
\author[COLSTATE,IPMU]{B.E.~Berger}
\author[LSU,atllnlnow]{S.~Dazeley}
\author[LBNL,IPMU,atniknow]{M.P.~Decowski}
\author[LBNL,atuwnow]{J.A.~Detwiler}
\author[UA,atargonnow]{Z.~Djurcic} 
\author[CALTECH,LBNL]{D.A.~Dwyer}
\author[UT]{Y.~Efremenko}
\author[RCNS,IPMU,atuwnow]{S.~Enomoto} 
\author[LBNL,IPMU]{S.J.~Freedman\fnref{deceased}}
\author[LBNL,IPMU]{B.K.~Fujikawa}
\author[RCNS]{K.~Furuno} 
\author[RCNS]{A.~Gando}
\author[RCNS]{Y.~Gando}
\author[STANFORD]{G.~Gratta}
\author[LSU,atjparcnow]{S.~Hatakeyama}
\author[WISC,atyalenow]{K.M.~Heeger}
\author[LBNL,atfnalnow]{L.~Hsu}
\author[RCNS,aticrrnow]{K.~Ichimura} 
\author[RCNS,IPMU]{K.~Inoue}
\author[RCNS,aticeppnow]{T.~Iwamoto}
\author[UT]{Y.~Kamyshkov}
\author[TUNL]{H.J.~Karwowski}
\author[RCNS,IPMU]{M.~Koga}
\author[RCNS,IPMU]{A.~Kozlov}
\author[DREXEL]{C.E.~Lane}
\author[UH]{J.G.~Learned}
\author[DREXEL]{J.~Maricic}
\author[TUNL]{D.M.~Markoff}
\author[UH]{S.~Matsuno}
\author[KSU,atmssnow]{D.~McKee}
\author[CALTECH,atjlabnow]{R.D.~McKeown}
\author[DREXEL]{T.~Miletic}
\author[RCNS]{T.~Mitsui}
\author[RCNS]{M.~Motoki}
\author[RCNS,atkeknow]{K.~Nakajima} 
\author[RCNS,atrcnpnow]{K.~Nakajima} 
\author[RCNS,IPMU]{K.~Nakamura}
\author[LBNL]{T.~O'Donnell}
\author[RCNS,IPMU,aticrrnow]{H.~Ogawa}
\author[RCNS,atcennow]{F.~Piquemal}
\author[RCNS,atlpscnow]{J.-S.~Ricol}
\author[RCNS]{I.~Shimizu}
\author[RCNS]{F.~Suekane}
\author[RCNS]{A.~Suzuki}
\author[LSU,atucdnow]{R.~Svoboda}
\author[RCNS,atkeknow]{O.~Tajima}
\author[RCNS]{Y.~Takemoto}
\author[RCNS]{K.~Tamae}
\author[STANFORD]{K.~Tolich}
\author[TUNL]{W.~Tornow}
\author[RCNS]{Hideki~Watanabe}
\author[RCNS]{Hiroko~Watanabe}
\author[LBNL,atuclanow]{L.A.~Winslow}
\author[RCNS,atosakanow]{S.~Yoshida}
%
%
%
\address[UA]{Department of Physics and Astronomy, University of Alabama, Tuscaloosa, Alabama 35487, USA}
\address[RCNS]{Research Center for Neutrino Science, Tohoku University, Sendai 980-8578, Japan}
\address[CALTECH]{W.~K.~Kellogg Radiation Laboratory, California Institute of Technology, Pasadena, California 91125, USA}
\address[UTChem]{Department of Chemistry, University of Tennessee, Knoxville, Tennessee 37996, USA}
\address[UH]{Department of Physics and Astronomy, University of Hawaii at Manoa, Honolulu, Hawaii 96822, USA}
\address[COLSTATE]{Department of Physics, Colorado  State University, Fort Collins, Colorado 80523, USA}
\address[LSU]{Department of Physics and Astronomy, Louisiana State University, Baton Rouge, Louisiana 70803, USA}
\address[LBNL]{Physics Department, University of  California, Berkeley and Lawrence Berkeley National Laboratory,  Berkeley, California 94720, USA}
\address[UT]{Department of Physics and Astronomy, University of Tennessee, Knoxville, Tennessee 37996, USA}
\address[STANFORD]{Physics Department, Stanford University, Stanford, California 94305, USA}
\address[WISC]{Department of Physics, University of Wisconsin at Madison, Madison, Wisconsin 53706, USA}
\address[KSU]{Department of Physics, Kansas State University, Manhattan, Kansas 66506, USA}
\address[TUNL]{Triangle Universities Nuclear Laboratory, Durham, North Carolina 27708, USA and
Physics Departments at Duke University, North Carolina State University, and the University of North Carolina at Chapel Hill, North Carolina, USA}
\address[DREXEL]{Physics Department, Drexel University, Philadelphia, Pennsylvania 19104, USA}
\address[IPMU]{Kavli Institute for the Physics and Mathematics of the Universe (WPI), University of Tokyo, Kashiwa 277-8583, Japan}
%
%
%
\address[atllnlnow]{Present address: Lawrence Livermore National Laboratory, Livermore, California 94550, USA}
\address[atucdnow]{Present address: Department of Physics, University of California Davis, Davis, CA 95616, USA}
\address[aticrrnow]{Present address: Kamioka Observatory, Institute for Cosmic Ray Research, University of Tokyo, Gifu 506-1205, Japan}
\address[atokayamanow]{Present address: Department of Physics, Okayama University, Okayama 700-8530, Japan}
\address[atlanlnow]{Present address: Los Alamos National Laboratory, Los Alamos, New Mexico 87545, USA}
\address[atbnlnow]{Present address: Physics Department, Brookhaven National Laboratory, Upton, New York 11973, USA }
\address[atniknow]{Present address: Nikhef and the University of Amsterdam, Science Park, Amsterdam, Netherlands}
\address[atuwnow]{Present address: Center for Experimental Nuclear Physics and Astrophysics, University of Washington, Seattle, Washington 98195, USA}
\address[atargonnow]{Present address: Argonne National Laboratory, Argonne, Illinois 60439, USA}
\address[atjparcnow]{Present address: J-PARC Center, 2-4 Shirane Shirakata, Tokai-mura, Naka-gun, Ibaraki 319-1195, Japan}
\address[atyalenow]{Present address: Department of Physics, Yale University, New Haven, Connecticut 06520, USA}
\address[atfnalnow]{Present address: Fermi National Accelerator Laboratory, Batavia, Illinois 60510, USA}
\address[aticeppnow]{Present address: ICEPP, University of Tokyo 7-3-1 Hongo, Bunkyo-ku, Tokyo 113-0033, Japan}
\address[atosakanow]{Present address: Graduate School of Science, Osaka University, Toyonaka 560-0043, Japan}
\address[atmssnow]{Present address: Department of Chemical and Physical Sciences, Missouri Southern State University, Joplin, Missouri 64801, USA}
\address[atjlabnow]{Present address: Thomas Jefferson National Accelerator Facility, Newport News, Virginia 23606, USA}
\address[atrcnpnow]{Present address: Research Center for Nuclear Physics, Osaka University, Ibaraki 567-0047, Japan}
\address[atcennow]{Present address: CEN Bordeaux-Gradignan, CNRS/IN2P3 and Universit\'{e} de Bordeaux I, F-33175 Gradignan Cedex, France}
\address[atlpscnow]{Present address: LPSC Universit\'{e} Joseph Fourier, CNRS/IN2P3, Institut Polytechnique de Grenoble, Grenoble, France}
\address[atkeknow]{Present address: High Energy Accelerator Research Organization, KEK, Tsukuba 305-0801, Japan}
\address[atuclanow]{Present address: Department of Physics and Astronomy, University of California Los Angeles, Los Angeles, CA 90095, USA}
\fntext[deceased]{Deceased.}

%% file: Intro.tex
\section{Introduction}
\label{Sec:Intro}

\par Sensitive detection of trace radioactivity is important for 
low energy solar neutrino, dark matter, and neutrinoless double beta decay experiments.  
Due to demanding constraints on detector backgrounds it has become essential to better 
understand the mechanisms that can be used to remove trace-level radioactivity from detector 
components, specifically from the active volume.   This study was motivated by the goal of 
expanding the physics capability of the KamLAND experiment to include detection of low energy 
solar neutrinos and other events from rare physics phenomena.

\par The KamLAND detector is situated in the Kamioka mine, Gifu prefecture, Japan and 
was commissioned in 2002 to test the large mixing angle (LMA) solution of the solar 
neutrino problem~\cite{bib:BahcallSolarTheory,bib:SolarNeutrinoProblem}.  KamLAND 
demonstrated the validity of the LMA solution by observing the disappearance of 
\antinu s produced in nuclear reactors in 
Japan~\cite{bib:KamLAND_PRL_1,bib:KamLAND_PRL_2,bib:KamLAND_PRL_3,bib:KamLAND_PRD_4}.  
KamLAND's active target is 1 kton of organic liquid scintillator (\ls), composed of 
80.2\% $n$-dodecane, 19.8\% 1,2,4-trimethylbenzene (PC) and~$1.36 \pm 0.03$ g/L of 
2,5-diphenyloxazole (PPO).  

\par
\begin{figure}[t!]
\centering 
\includegraphics[keepaspectratio,width=0.45\textwidth,trim= 15mm 5mm 25mm 0mm]{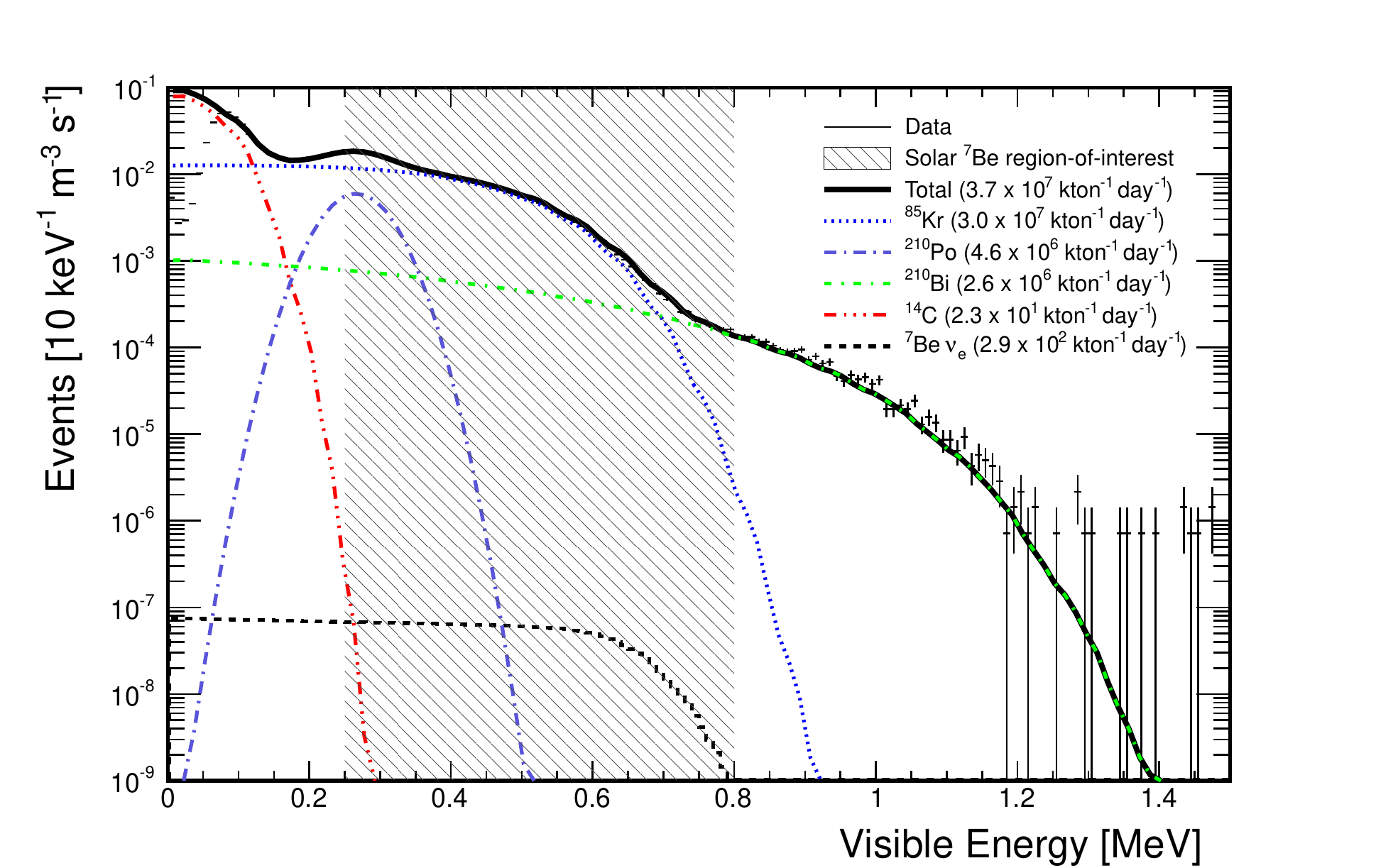}   
\caption[]{KamLAND low energy singles spectrum. The data is displayed together 
with a background model. The major background components are indicated along with 
their integral rates inside the energy interval~$\left[0.25,0.8\right]$ MeV (hatched 
region).  The expected \BeNeutrino\ solar neutrino recoil event spectrum (calculated using BS05 solar 
rates~\cite{bib:BS05OP})  is overlaid to provide the appropriate scale. All spectra 
are converted to visible energy with KamLAND's energy response function and
folded with a $\sigma = 7.8 \%/ \sqrt{ E~[{\rm MeV}]}$ 
Gaussian energy resolution~\cite{bib:KeeferThesis}.}
\label{Fig:SolarRegionWithBackgrounds}
\end{figure}

  The statistical analysis of KamLAND's low energy event distribution is performed 
by fitting summed beta spectra of different radionuclides to KamLAND's singles
data (events not requiring delayed coincidence) and
allowing their normalizations to float freely. 
The result of this procedure is
depicted in Fig.~\ref{Fig:SolarRegionWithBackgrounds} together with the data.
The mass concentrations of the contaminants appearing in
Fig.~\ref{Fig:SolarRegionWithBackgrounds} are listed along with several others
in Table~\ref{tab:KamLANDConc}~\cite{bib:KeeferThesis}.
Analysis of the singles data yields an integral event rate 
of~$3.7 \times 10^{7}$~kton$^{-1}$~day$^{-1}$ in the energy window of interest ($\left[0.25,0.8\right]$ MeV). 
This rate is to be compared to the expected \BeNeutrino\ solar neutrino elastic
scattering signal of
291~events~kton$^{-1}$~day$^{-1}$. 
Spectral analysis allows the identification of the principal contributors to KamLAND's 
background to be \krypton, $^{210}$Bi and $^{210}$Po, the latter two supported
by the decay of long-lived $^{210}$Pb~\cite{bib:KeeferThesis}. 
\krypton\ is a fission product, found in the atmosphere mainly due to nuclear fuel 
re-processing~\cite{bib:TraceNobleGases}, while $^{210}$Bi and $^{210}$Po are \radon\ decay daughters.
As both \krypton\ and \radon\ are airborne, it was assumed that these contaminants were introduced 
into the KamLAND \ls\ during filling from exposure to air.
The initial growth of $^{210}$Po, observed in early KamLAND data,
supports this hypothesis~\cite{bib:DjurcicThesis}. After the initial growth the $^{210}$Po
activity leveled off, indicating that it was supported
by its long-lived parent \lead. The analysis of delayed $^{214}$Bi-$^{214}$Po
beta-alpha coincidences, which occur above \lead\ in the U-decay chain,
provided an effective $^{238}$U
decay rate of $3.7\pm 0.5$ kton$^{-1}$ day$^{-1}$. This excludes the possibility 
that $^{210}$Po, observed with an event rate of $4.6\times 10^6$ kton$^{-1}$ day$^{-1}$,  
was supported by either $^{238}$U or $^{226}$Ra.
Although there was no direct spectroscopic evidence for the presence of 
\argon activity (an atmospheric spallation product), its existence was 
suspected because of the air exposure hypothesis, and a limit on its
concentration is
included for completeness in Table~\ref{tab:KamLANDConc}.

\input{Table01}

\par The studies presented in this paper were undertaken to understand
if and how the internal background rate,
five orders of magnitude larger than the solar neutrino signal,
can be sufficiently reduced by means of chemical purification.   
At the fundamental level, removal of impurities 
can be achieved by differences in chemical potentials between different media and phases.  
The diffusion of impurities from one medium to another is required to achieve 
equilibrium~\cite{bib:Crank}.  
Consequently we studied the distribution of impurities identified to be responsible
for KamLAND's low energy background with regards to their behavior in
different media and phases.  These included:
\begin{itemize}
\item vapor-liquid phases in distillation
\item solid-liquid phases in adsorption, as well as filtering and isotope
exchange
\item liquid-liquid phases in water extraction.
\end{itemize}
In addition we studied the impact of heating on the \Lead\ removal efficiency to target possible 
organometallic \Lead\ compounds formed after alpha decay and known to be prone to heat-induced break up. This 
paper reports measured reduction factors for these different purification methods.
Studies done on the removal of radioactive gases (\krypton\ and \radon) are described in ~\cite{bib:TakeuchiMastersThesis} 
and are not covered in this work.
Final reduction factors achieved from full-scale scintillator purification in the KamLAND
detector are not within the scope of this paper.

\par The experimental procedures used throughout this study are outlined in Sec.~\ref{Sec:ExpMethods}.  
In Sec.~\ref{Sec:RnDaughterRemoval} we discuss the analytical techniques and methods 
used to quantify and remove \Radon\ decay products from~\ls\ and its
constituents  PC, $n$-dodecane, and PPO. Finally, in 
Sec.~\ref{Sec:Conclusion}, we discuss the implications and significant findings of this work.

%% file: Table01.tex
\renewcommand\arraystretch{1.25}
%
\begin{table}[bdp]
  \caption[]{Measured radioactivity concentrations in KamLAND~\ls.  The limit for $^{39}$Ar was derived from solubility arguments~\cite{bib:KeeferThesis}. }
  \begin{center}
    \begin{tabular*}{\columnwidth}{@{\extracolsep{\fill}} l l c }
      \hline
      \textbf {Isotope}  &  \textbf {Concentrations [g/g]}       \\ \hline \hline
      $\rm ^{39}Ar $         & $\rm < 4.3 \times10^{-21}        $               \\ 
      $\rm ^{40}K $          & $\rm (1.30 \pm 0.11)\times10^{-16} $           \\ 
      $\rm ^{85}Kr $         & $\rm (6.10 \pm 0.14)\times10^{-20} $           \\ 
      $\rm ^{210}Pb$         & $\rm (2.06 \pm 0.04)\times10^{-20}   $           \\ 
      $\rm ^{232}Th$         & $\rm (8.24 \pm 0.49)\times10^{-17} $           \\ 
      $\rm ^{238}U$          & $\rm (1.87 \pm 0.10)\times10^{-18} $            \\ \hline
    \end{tabular*}
  \label{tab:KamLANDConc}
  \end{center}
\end{table}
%

%% file: Procedure.tex
\section{Experimental Procedures and Detection Methods}
\label{Sec:ExpMethods}

\par Measuring the large reductions in \lead\ needed for KamLAND's
low background phase required the quantitative determination of \Pb\ concentrations 
at a level of $10^{-24}$ to $10^{-20}$~g/g. The starting value
corresponds to approximately $4.6\times 10^{4}$ \lead\ atoms per liter of \ls,
while the smaller concentration represents the goal for 
adequately purified \ls.  Clearly this is a challenging analytical
problem. 

\par The measurement of \Pb\ concentrations at such extremely low levels 
is best achieved using radioactive tracers. The isotope of concern,
\lead, is a low Q-value beta emitter. Its daughters, $^{210}$Bi and
$^{210}$Po, are pure beta and alpha emitters, respectively. None of these
decays offer a convenient experimental signature suited to measure small
concentrations in lab scale experiments. Furthermore, \lead\ has
a half life of 22.3~y resulting in a relatively
small specific activity of $4.5\times 10^{-5}$~Bq/L at the starting
concentration of $10^{-20}$~g/g.
To overcome this experimental challenge, we 
assume that all isotopes of \Pb\ born via alpha decay
have similar chemical characteristics in organic media, 
and thus, similar responses to purification techniques.
Isotope-specific studies showed a factor of 3 higher \ThPb\ reduction than observed for \UPb
~\cite{bib:KeeferThesis}. This might be attributed to differences 
in nuclear recoil energies and \Pb\ residence times in the \ls.  Given the interest 
in large reduction factors, and not analytic precision, the possible impact of these effects are neglected in this work.

\par The short lived isotopes of $^{212}$Pb ($T_{1/2}=10.64$~h), 
born via $^{216}$Po ($^{220}$Rn daughter), and $^{214}$Pb ($T_{1/2}=26.8$~min), 
born via $^{218}$Po ($^{222}$Rn daughter),
can be used to study \Pb\ removal within our target concentration range of 
$10^{-20}$ to $10^{-24}$~g/g. 
At a concentration of $10^{-20}$~g/g the specific 
$^{212}$Pb activity is 0.83~Bq/L, a factor $1.8\times 10^4$
larger than that of \lead\ at equal concentration.  
Characteristic gamma radiation, of energies 238~keV and 351~keV, are emitted
after the beta decays
of $^{212}$Pb and $^{214}$Pb, respectively. This allows for convenient detection with a 
low-background germanium (Ge) detector. The subsequent fast Bi-Po beta-alpha decay sequences
following these \Pb\ decays allow the further utilization of delayed 
coincidence counting with \ls\ for high-sensitivity measurements.

\subsection{Determination of the Radio-nuclide Reduction Factors}
\label{RnSoures}

\par The effectiveness of various \Pb\ removal techniques was tested
by loading \ls\ with trace amounts of radioactive $^{212}$Pb or $^{214}$Pb.
This was achieved by first dissolving $^{220}$Rn or $^{222}$Rn,
exploiting the excellent solubility of \Rn\ in organic solvents,
and then allowing it to decay into the meta stable \Pb\ isotopes.
The dissolved \Pb\ activity, $\it{A_{i}} $, was measured after loading.
The remaining activity, $\it{A_{f}}$, was measured after the loaded solution had been 
subjected to a specific purification procedure. Both activities were
corrected to correspond to a common reference time.
The reduction factor $R$ for a given species is defined
as:
\begin{equation}
\it{ R \; =\; \frac{A_i}{A_f}}
\label{eq:eff}
\end{equation}

\par Two detector types were used to determine the \Pb\ activities.  
For samples with specific activities of more than 1~Bq/L, 
Ge detectors were utilized to measure 
the gamma radiation emitted in $^{214}$Pb, $^{214}$Bi, $^{212}$Pb, 
$^{212}$Bi and $^{208}$Tl beta decays. 
For low activity samples, liquid scintillation counting was employed, utilizing
the delayed beta-alpha coincidences from the 
$^{212}$Bi-$^{212}$Po ($^{220}$Rn decay chain) and $^{214}$Bi-$^{214}$Po
($^{222}$Rn decay chain) decay sequences.  The latter technique allowed the
nearly background-free measurement of specific activities down to 10~mBq/L.
Together, these detection techniques provided a means to perform studies of 
purification reduction factors spanning five orders of magnitude 
and reaching $^{212}$Pb concentrations as low as $10^{-22}$~g/g. Further sensitivity to
the target concentration of $10^{-24}$~g/g can only be achieved using the KamLAND
detector itself.

\subsection{Scintillator Loading and Purification Procedures for Radon Daughters}
\label{Sec:PurProcedure}

\par Gas flow-through sources from Pylon Electronics Inc. (models TH-1025 and RN-1025) 
were used to load \ls\ with $^{220}$Rn or $^{222}$Rn, respectively. A second
$^{220}$Rn source was
made using a ThO$_2$ powder.  In all instances N$_{2}$ was utilized as 
the carrier gas. To remove dust particles from the gas stream, the N$_2$ was
passed through 100~$\mu$m and 
0.8~$\mu$m filters prior to a flow-through source.
The \Rn-loaded N$_{2}$ gas was then passed through 500~cm$^{3}$ of \ls\ via a  
glass bubbler, fitted with a sparger to break up the gas for increased \Rn\ absorption.
Flow meters were attached to the input and output gas stream to monitor flow rates.  

\par Our studies showed that the use of a filtered carrier gas was important for achieving 
reproducible results. Particulates carried by the gas stream readily collect charged \Rn\ 
decay products and are efficiently removed by filtering, thus ``faking'' a
removal mechanism not present in KamLAND scintillator.

\par Lead removal experiments were performed with \ls,  $n$-dodecane, PC, and PPO.  
All samples were handled using the following standardized procedure (unless otherwise 
noted):  

\begin{enumerate}
\item  Sparge liquid with \Rn-loaded \nitrogen\ for a maximum of 24 hours.\footnote{
Care must be taken when loading \ls\ or pure PC.  PC has a high 
vapor pressure causing it to evaporate during bubbling, resulting in a 
lowering of the PC concentration as well as corrosion of filters and membranes.} 
\label{En:Bubbling}

\item  Transfer the $^{212,214}$Pb-loaded liquid into a counting container
to determine the initial activity $\it{A_{i}}$.
For counting with Ge detectors, two 125~cm$^{3}$
Nalgene$^{\textrm{\textregistered}}$ bottles were used.  
If liquid scintillation counting was to be used, half of the sample was 
transferred to a 10.80~cm diameter, 4.44~cm long acrylic  
cell, fitted with a 7.62~cm diameter photo-multiplier tube (PMT). 
The counted \ls\ volume was typically 200~cm$^{3}$. 
 
\item  When using the Ge detectors, the counted liquid was transferred into the 
purification system. When using liquid scintillation counting,
sufficient liquid was prepared to
allow parallel counting and purification.  This method assumes 
a homogeneous distribution of \Pb\ in the \ls.  \label{En:BottleCount}

\item  After purification, the liquid was counted using the appropriate 
detector to determine the final activity $A{_{f}}$.  \label{En:PurCount}

\end{enumerate}
\vskip 0.1in

\par  The ``before'' and ``after'' counting was always performed using the same detector 
to allow for cancellation of systematic errors.  In some instances it was necessary to use 
only a single component of the \ls\ during the purification process. If an \ls\ component was analyzed
by liquid scintillation counting a scintillating cocktail had to be 
made prior to loading the counting cell. This dilution was taken into account 
in the reported reduction factor.  Furthermore, measurements were carried out to cross 
calibrate the two different detector types.

\par Detailed studies of systematic errors were performed on every aspect of the 
purification processes.  Thorough cleaning of all containers used in the loading and
counting was imperative when dealing with extremely low levels 
of radioactivity. Cross contamination had to be avoided in order to achieve 
reproducible results.  
When chemically compatible, all components were cleaned with high purity solvents (analytic grade or better).  
Each part was cleaned with acetone, then with 1 molar HNO$_3$, followed by a rinsing with de-ionized (DI) water and ethanol.  

\par In the case of Ge detector counting, it was found that a substantial fraction
(up to 50\%) of the \Pb\ activity contained in the \ls\ could adhere to the
walls of the Nalgene bottles used for the counting. The bottles were
thus counted again after the \Pb\ loaded scintillator had been transferred
into the purification setup to account for this loss mechanism.

%% file: Results.tex
\section{Removal of Radon-Born Lead from Liquid Scintillator}
\label{Sec:RnDaughterRemoval}

\par In any purification process, the difference in concentrations and solubilities 
between phases drives the separation of impurities between them.  In this work
these phases were realized in the form of water-oil 
and silica-oil systems or gas-liquid phases.  Processes which rely primarily on diffusion of the 
impurities between phases require initial concentrations in the transient phase to be less than those 
listed in Table~\ref{tab:KamLANDConc}.  
The mass transfer involved in such diffusive processes can be modeled assuming thermodynamic 
equilibrium. The work presented here does not report the reduction in terms of a Gibbs potential, 
as done in previous publications~\cite{bib:Borexino-NIM-SilicaGel}. Instead, 
we report the results in the form of a reduction factor as defined in Eq.~\ref{eq:eff}. 
We found this to be sufficient for the purpose of identifying the most effective method of purification.

\par The following purification techniques were studied: water extraction, 
ion enhanced water extraction, isotope exchange, filtering, adsorption and distillation.  When 
applicable, experiments were performed in series with one another to establish if the methods 
scale as the product of their individual reduction factors. For each of these techniques, the reduction factor,
time of operation, solute volume, and initial activity were studied for correlations.  

\par Among the \radon\ daughters \lead\ is of particular interest because of 
its long half life.   
Following the assumption discussed in Section~\ref{Sec:ExpMethods} that different
alpha-born isotopes of lead in organic media have similar characteristics,
the purification techniques described below were evaluated using \ThPb\ and \UPb\ as tracers. 
However, one cannot assume that all  metallic \Radon\ daughters are removed with 
equal efficiency as they have different chemical characteristics.  
The radionuclides $^{208}$Tl, $^{212}$Bi, $^{214}$Bi and $^{218}$Po were analyzed 
separately to better understand the variation in the reduction factors across
chemical species. For example, experimental information obtained 
on $^{218}$Po was used to gain some insight into how much $^{210}$Po ($T_{1/2}=138\; \rm d$)
would remain in the \ls\ after purification.  The results of these non-\Lead\ purification studies are not reported here.
More details on the purification research performed with KamLAND \ls\
can be found in~\cite{bib:KeeferThesis,bib:KishimotoAIPProc,bib:TakeuchiMastersThesis,bib:KosekiMastersThesis,bib:EbiharaMastersThesis,bib:KibeMastersThesis,bib:MomoyoMastersThesis,bib:KeeferNuclPhysBProc,bib:KeeferAIPProc}.


\subsection{Water Extraction}
\label{WaterExt}

\par A water purification system already existed in KamLAND; it had
been used during the initial filling~\cite{bib:OgawaThesis}. We thus first
tried to understand whether this system could be used, perhaps with
modification, to remove dissolved \Lead\ from KamLAND \ls.
A miniature water extraction system was built to circulate \ls\ through a glass 
purification column filled with DI water.  The \ls, which is less 
dense than water, was removed from the top of the water column and re-circulated 
back to the bottom of the column using a peristaltic pump.  To increase the surface 
to volume ratio, the \ls\ was broken into small bubbles with a custom-made glass 
frit affixed to the bottom of the column.

\par  Water extraction experiments showed weak \Lead\ reduction factors between
1.02 and 1.10. Using about 500 cm$^3$ of water, 250 cm$^3$ of \ls\ and a
flow rate of 200 $\rm cm^3/min$, the observed reduction factors saturated for circulation times longer than 5 hours~\cite{bib:KeeferThesis}.
A second set of independent experiments was performed using 50 cm$^3$ of \Lead-loaded \ls\ with 60 cm$^3$ of 
distilled water mixed with 0.01 to 6 mol/L of HNO$_3$, HCl, NaOH~\cite{bib:KosekiMastersThesis}. 
Mixing was achieved by vigorously shaking for 5 min.  
In these experiments we observed saturation levels consistent with the more controlled mixing noted above.
In all experiments the \Lead\ reduction factor showed a clear decrease with growing pH.  
However, the magnitude of this change over the pH range never exceeded 30\%. 
Therefore, water extraction was found to be not well suited for achieving the
very large \Lead\ reduction factors needed in KamLAND.


\subsection{Isotope Exchange, Filtration, and HTiO Adsorption}
\label{Sec:IsoFiltering}

\par \ls\ loaded with radioactive \Lead\ was circulated through a bed of 
granular \Lead, or \Lead\ shot. Assuming that the dissolved \Lead\ forms
an equilibrium with the essentially infinite number of stable \Lead\
atoms, we tested the possibility of isotope exchange. These experiments 
showed no quantifiable \Lead\ reduction effect beyond the typical loss 
to the surfaces of the pumping system.

\par Early studies showed that filtering loaded \ls\ removed the 
dissolved \Lead\ in a variable manner. This fluctuation in lead removal efficiency was traced to 
the presence of dust particles in the carrier gas stream to which the charged radon decay products readily attach~\cite{bib:Migdal}. 
Carefully filtering the gas stream avoided this interference and enhanced the reproducibility 
of the experiments. 

\par HTiO has successfully been used in the SNO experiment to remove actinides 
from light and heavy water~\cite{bib:SNO_HTIO}.  
We investigated this method for its suitability in removing \Lead\ from organic
\ls. HTiO pre-loaded HEPA PTFE filters (surface density 0.64 g/cm$^{2}$),
supplied by the Oxford University SNO group, were utilized for these tests.
\Lead-loaded \ls\ was placed into a 60 ml syringe attached to the filter and the
liquid was pushed slowly through the filter. The filter was replaced after every 125 ml 
of \Lead-loaded \ls. For flow rates of 3.3 and 16.5~cm$^3$/min we observed
\Lead\ reduction factors of $1.06 \pm 0.01$ and $1.07 \pm 0.01$, respectively.
No further consideration was given to this method.


\begin{figure}[tbp]
\begin{center}
  \subfigure[]{\includegraphics[keepaspectratio,height=0.65\columnwidth]{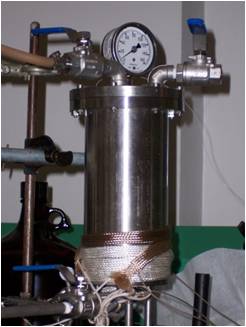}\label{Fig:HeatingApp}}
  \subfigure[]{\includegraphics[keepaspectratio,height=0.65\columnwidth]{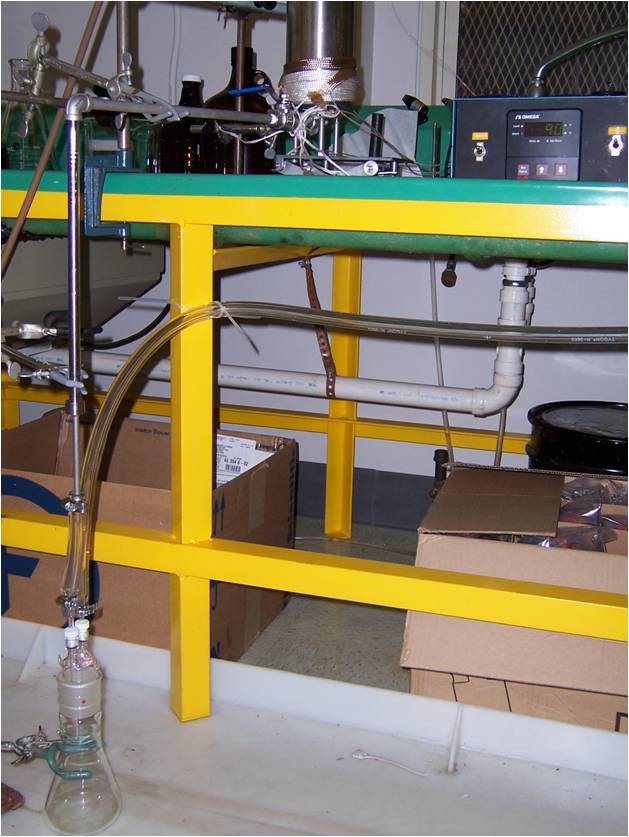}\label{Fig:AdsorbApp}}
\caption{The stainless steel vessel used to heat the \ls\ prior to adsorption is depicted in Fig.~\ref{Fig:HeatingApp}. Fig.~\ref{Fig:AdsorbApp} is the adsorption column in which all silica gel experiments were performed.  At the top one can see the heating vessel and thermocouple.  Furthermore, a vacuum line was used to regulate the flow rate through the silica column.}
\end{center}
\end{figure}

\subsection{Chromatography with Silica Gels (Adsorption)}
\label{Sec:AdsorbentPur}

\par Previous studies identified silica gel extraction as an
efficient means of removing inorganic impurities from \ls ~\cite{bib:Borexino-NIM-SilicaGel}.
This method was studied in depth to evaluate its suitability for removing \Lead\ 
using different adsorption materials packed in chromatographic columns. 
Plate theory~\cite{bib:UngerKK} is commonly used to describe liquid-solid 
chromatography, which requires mass and thermodynamic equilibrium between the 
mobile (\ls) and stationary (silica gel) phases.  One can
divide the column into many distinct sections, referred to as plates.
The concentration of an impurity in successive plates can be derived from mass balance and 
equilibrium conditions:
\begin{equation}
\rm X_{M(p)} \; =\; \frac{X_{0}\nu^{p}}{p!} e^{-\nu},  \;\;\;\;\;  p=1,2,3,...
\label{eq:ElutionCurve}
\end{equation}
where $\rm X_{0}$ is the initial concentration of the impurity, $\rm X_{M(p)}$ is the concentration 
in the p$\rm ^{th}$ plate, and $\nu$ is the plate volume.  Eq.~\ref{eq:ElutionCurve} is referred 
to as the elution curve and
indicates that given enough adsorbent, there is no theoretical limit to the achievable reduction factor.

\par There are two possible methods for purifying \ls\ with this technique.  One can assume that the silica gel
is infinitely adsorptive and run the liquid through a column in a loop mode, or
make a single pass through the column in a batch mode.
Experiments showed batch purification through a long column to be the method of 
choice, as multiple passes through the same column lead to lower reduction
factors.  Batch mode makes efficient use of the concentration gradient of the 
solute developing along the column.
Our observation that batch mode provides higher purification factors 
than loop mode agrees well with previous studies~\cite{bib:Borexino-NIM-SilicaGel}.
To illustrate this point, Table~\ref{Ta:BatchLoop} lists the \Lead\ reduction 
factors obtained for three different types of silica gel used in both batch 
and loop mode.

\input{Table02}

\par All results presented from here on are based on batch mode operation.  
A wide variety of silica gels are commercially available. For this study
a $32-63 \; \mu$m (grain size) silica gel from Selecto Scientific was used as the standard.  
Our laboratory chromatographic \Lead\ purification system, shown in Fig.~\ref{Fig:AdsorbApp},
was designed and operated in the following way: a slightly pressurized stainless 
steel reservoir, holding approximately 10 L of \ls, was coupled
to a 35~cm long stainless steel chromatographic column with 0.97~cm inner diameter.
Valves installed in the \ls\ 
path allowed flow regulation through the column.  The 
column was fitted with a $10 ~\mu$m stainless steel mesh filter at the 
outlet to sequester the adsorbent.  

\input{Table03}

\par  Prior to runs with \ls, the silica gel was first conditioned (compacted). This was
done to keep eddies and voids from developing inside the column.
Sufficient compactification was achieved by providing constant \nitrogen\
pressure to the top of the column while pulling partial 
vacuum at the bottom (without \ls).   
Our standard purification procedure gave consistent and reproducible \Lead\ removal 
factors, as verified by repeated measurements with the standard gel. The procedure 
consisted of the following steps:
\vskip 0.1in
\begin{enumerate}
  \item Clean all parts with acetone, then with 0.1 - 1.0 molar HNO$_{3}$, followed by a rinse with DI water and ethanol.  
  \item Add the desired amount of silica gel to the column, close all valves, and pump down
        to 100~hPa. Assist setting of the silica granules by knocking the column during pump-down.
  \item Apply 1000~hPa \nitrogen\ pressure to the top of the column and continue to knock the column 
        to set the silica in place.  
        The pressure at the outlet rises to approximately 300~hPa when
        equilibrium is reached.
  \item Remove the vacuum pump and slowly release the pressure from the inlet side of the system.  
  \item Add the \Lead-loaded \ls\ to the reservoir with all valves closed.  Pressurize the reservoir
        with \nitrogen\ to 1000~hPa.
  \item Slowly open the valve between the reservoir and the column to wet the gel. 
        Controlled release of \ls\ prevents cavities from forming at the top of the silica column. 
        The flow rate of \ls\ through the column was 50 ml/min.
  \item After purification, the \ls\ is passed through a 0.2~$\mu$m PTFE HEPA filter to remove 
         silica particulates suspended in the \ls.
\end{enumerate}
%
%

\par Further conditioning tests were performed by heating the silica gel to
100 - 120~\degC\ to remove adsorbed water from its 
surface.  Water molecules are known to adhere to the adsorption sites and reduce the 
effectiveness of the silica gel~\cite{bib:UngerKK}.  However, this conditioning was not
found to provide any benefit in terms of boosting \Lead\ removal.

\par Using the procedure outlined above, \Lead\ removal studies were performed with many different 
types of adsorbents. The results of these experiments are given in Table~\ref{Ta:AdsorbentEff}.  
Our procedure provided reproducible results, and 
typical statistical uncertainties on individual runs were of order 0.1\%.
Repeated runs with the $32-63$~$\mu$m silica gel yielded a standard deviation
of $7.2\%$ for the \Lead\ reduction factor, which we interpret as a measure of the systematic uncertainty.

\par Since, as already mentioned, the elution curve (Eq.~\ref{eq:ElutionCurve}) indicates that 
in principle the reduction factor has no limit, making the chromatographic column arbitrarily long should allow
one to achieve any desired reduction factor. This behavior was not observed, as a maximal reduction factor of 29 was 
achieved for a single pass purification.  While this factor is substantial, it
falls short of the required value for KamLAND, so
studies were performed to determine whether this limitation could be overcome.


\label{OrganoMettalicComp}

\par We considered three hypotheses to explain the observed limit in the \Lead\
reduction factor during silica gel extraction:

\begin{Hypothesis}\label{Hyp:CrossContam}
There exists \Lead\ cross-contamination between runs (insufficient cleaning).
\end{Hypothesis}

\begin{Hypothesis}\label{Hyp:SiContam}
The intrinsic uranium and thorium concentrations of the silica gel, and the resulting out-gassing \Radon, re-introduces 
\ThPb\ into the \ls. The observed limit may be determined by the equilibrium between the removal and addition of Pb. 
\end{Hypothesis}

\input{Table04}

\begin{Hypothesis}\label{Hyp:OrganContam}
Some fraction of the \ThPb\ may exist in the \ls\ in a form without affinity towards the 
silica gel.
\end{Hypothesis}

\par Blank runs performed with unloaded \ls\ following a purification run with lead-loaded \ls\
did not show any carry-over activities, thus excluding
Hypothesis~\ref{Hyp:CrossContam} and constraining also
Hypothesis~\ref{Hyp:SiContam}. Furthermore, bottles and
containers were not re-used whenever
possible, and all re-used equipment underwent vigorous cleaning, as described in Sec.~\ref{Sec:PurProcedure}.

\par Hypothesis~\ref{Hyp:SiContam} was further refuted by measuring the
radioactive content of the gel. 
The $32-63$~$\mu$m Selecto Scientific silica gel was measured 
to have an intrinsic uranium and thorium content of 167 ppb and 152 ppb, 
respectively~\cite{bib:EXO_Materials}, resulting in 1.7~mBq of \ThPb\ in 2.5~g of silica gel.  
Typical loading activities for the \ls\ were 5~kBq. This silica gel had an average reduction factor of 19, 
resulting in 250~Bq of residual \ThPb\ activity.  
The total intrinsic activity of the silica gel is thus much smaller than the
measured activity after silica gel purification, implying that the former
contributes negligibly to the observed limit in the reduction factor.

\par  Hypothesis~\ref{Hyp:OrganContam} was also tested experimentally.  Chromatographic impurity removal 
techniques selectively address polar molecules or ionic states~\cite{bib:UngerKK,bib:SurfaceConc-OH}.  
It is the excess charge on the molecule, or atom, which draws it to the hydroxyl group (--OH) on 
the active adsorption sites of the gel.  The excellent reproducibility of our \Lead\ removal experiments
suggests that about 3-5\% of the \Lead\ atoms were present in a non-polar, molecular state, 
without affinity to the active hydroxyl groups of silica gel.

\par In order to understand why this species has no affinity to the silica gel,
it is of particular interest to examine a study performed with $n$-dodecane exposed to a $\rm ^{60}Co$ 
source~\cite{bib:DodecaneRadiation}.  This study showed that the exposure to ionizing radiation produced 
fragmented carbon chains. We note that the \ThPb\ and \UPb\ isotopes used in our studies, 
as well as the \lead\ present in KamLAND's \ls, are born via alpha decay. 
A recoiling \Lead\ nucleus, with typical energies of a few hundred keV, can easily fragment the carbon-carbon 
bonds in $n$-dodecane, having typical binding energies of order 10 eV~\cite{bib:CRCChemPhys}.  These fragmented molecular states 
could form stable, electrically neutral, non-polar compounds. The most notable of these are the \CPb\ bonds which 
tend to be covalent, and thus non-polar.  The decay of the parent nuclei into \Lead\ occurs in a carbon rich environment
and this bonding is energetically favored over many other available chemical
bonds (e.g.~C--C bonds)~\cite{bib:CRCChemPhys}. 
The non-polar \CPb\ compounds would have no affinity to the polar reaction sites
on silica gel, supporting Hypothesis~\ref{Hyp:OrganContam}.

\par  Table~\ref{Tab:OrganoPbProp} lists the relevant properties of the known
organic compounds containing \Lead. All exhibit lower boiling points than those of PC and $n$-dodecane.  
It is of interest to note $\rm C_{8}H_{20}Pb$ is thermally unstable and will decompose at 200~\degC.  
This process would leave behind a polar or ionic \Lead\ species.

\par To obtain supporting evidence for Hypothesis~\ref{Hyp:OrganContam}, we experimented with 
chemical procedures targeting the organic \CPb\ bonds.
It is known that the reagents: $\rm FeCl{_3}$~\cite{bib:FeCl3}, $\rm SnCl_{4}$~\cite{bib:TinCl4}, 
$\rm MoS_{2}$~\cite{bib:MOS} and thiol resin~\cite{bib:ThiolResin} target these \CPb\ bonds directly.

\par To test the effect of these chemicals on the \Lead\ reduction factor, we
first passed \ThPb-loaded \ls\ 
through a silica gel column to remove the ionic \Lead\ component. In a second step, 150 ml of the
treated \ls\ was stirred magnetically, with one of
the reagents listed above, for 45 min. Next, 15 ml of 0.5 M EDTA ($\rm [CH_2N(CH_2CO_2H)_2]_2$)
solution (pH of 8) was added to the mixture and stirred for an additional 15 min.  The mixture was filtered 
under partial vacuum and passed over a molecular sieve to remove any water.  The resulting \ls\ was 
filtered a second time to remove any particulates of the molecular sieve. The liquid was
counted to determine the remaining amount of \ThPb.
The \Lead\ reduction factors obtained this way are tabulated in Table~\ref{tab:OrganicData}.

\input{Table05}

\par \ls\ treated with $\rm FeCl{_3}$ before silica gel extraction showed a
50-fold enhanced \Lead\ reduction factor compared to plain chromatography.
We interpret this boost in effectiveness as supporting evidence for Hypothesis~\ref{Hyp:OrganContam}.
However, the $\rm FeCl{_3}$ treatment of the \ls\ was found to permanently degrade its optical properties.  
The measured light attenuation length at 436 nm was $< 10^{-4}$ m compared to approximately 13 m for
untreated \ls.  Optical transparency could not be restored by means of filtration or repeated
treatment with silica gel.

\par The successful experiments utilizing reagents targeting \CPb\ bonds
prompted us to consider heating the \Lead-loaded \ls\ to 100-200~\degC\
to remove the tetraethyllead which decomposes~\cite{bib:CRCChemPhys}.
As a test, \ThPb-loaded \ls\ was pre-heated in a stainless steel vessel, as shown in Fig.~\ref{Fig:HeatingApp}, before sending it
through the ion-exchange column, depicted in Fig.~\ref{Fig:AdsorbApp}. The results from these heating experiments are provided
in Table~\ref{tab:HeatingTable}.  The first three entries show a substantial
enhancement of the \Lead\ reduction factor for pre-heated \ls\ compared to the
experiments with silica gel extraction only.
The following three entries give a factor 5 to 10 reduction obtained by just heating the \ls\ and
not using silica gel.
Lead adherence to the container walls was excluded by experiments performed at room temperature. 

\input{Table06}

\par We consider these results to be further evidence for 
 Hypothesis~\ref{Hyp:OrganContam}.  However, the data are not specific enough as to allow us to 
identify which compounds are present and in what proportion. This is nevertheless an important 
finding, indicating that the limitless reduction factors predicted by plate theory,  Eq.~\ref{eq:ElutionCurve},
are not to be expected for alpha decay products in an organic medium.


\begin{figure*}[tbp]
\begin{center}
  \subfigure[]{\includegraphics[height=0.73\columnwidth, width=0.9\columnwidth]{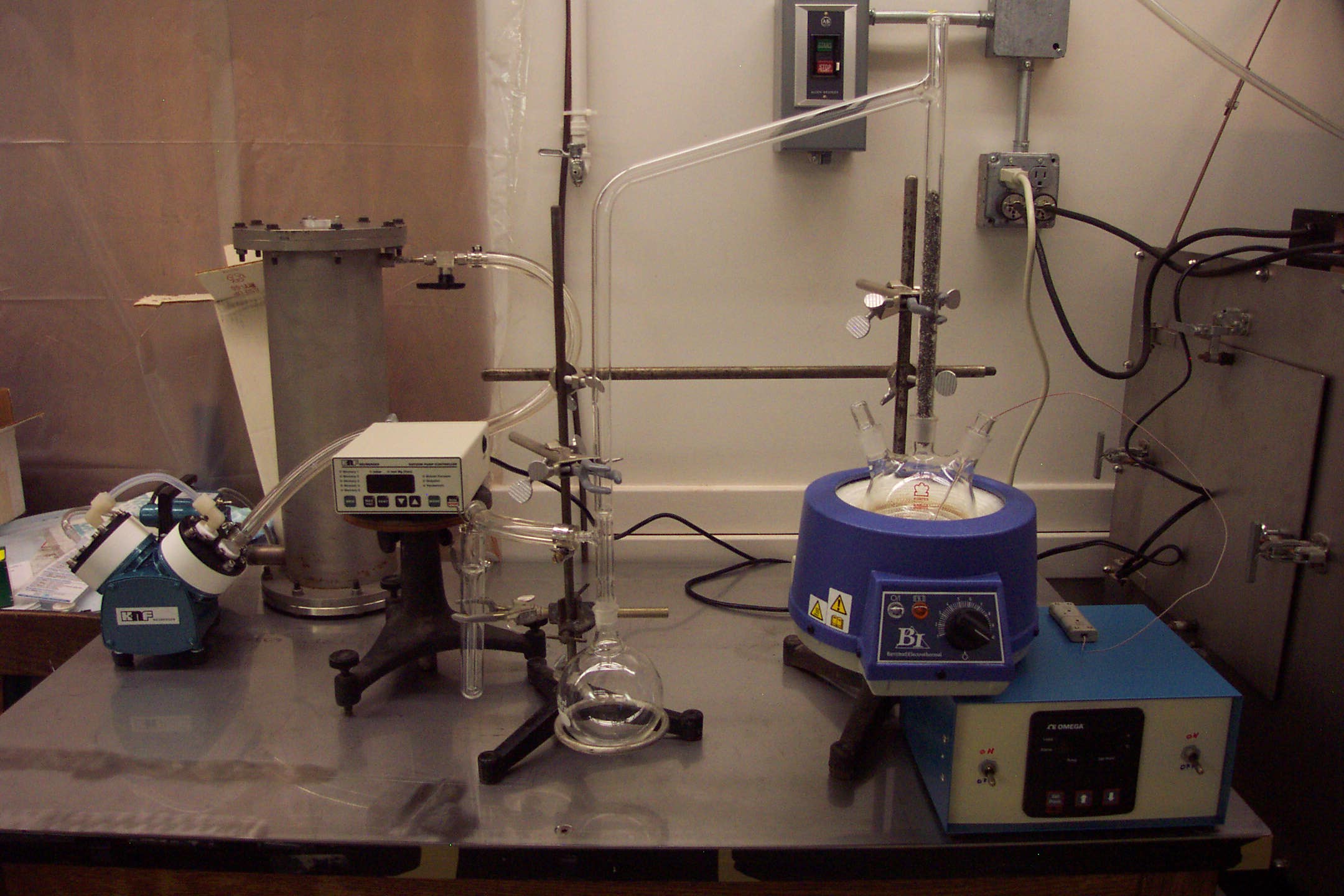}\label{Fig:UADistillation}}
  \subfigure[]{\includegraphics[height=0.73\columnwidth]{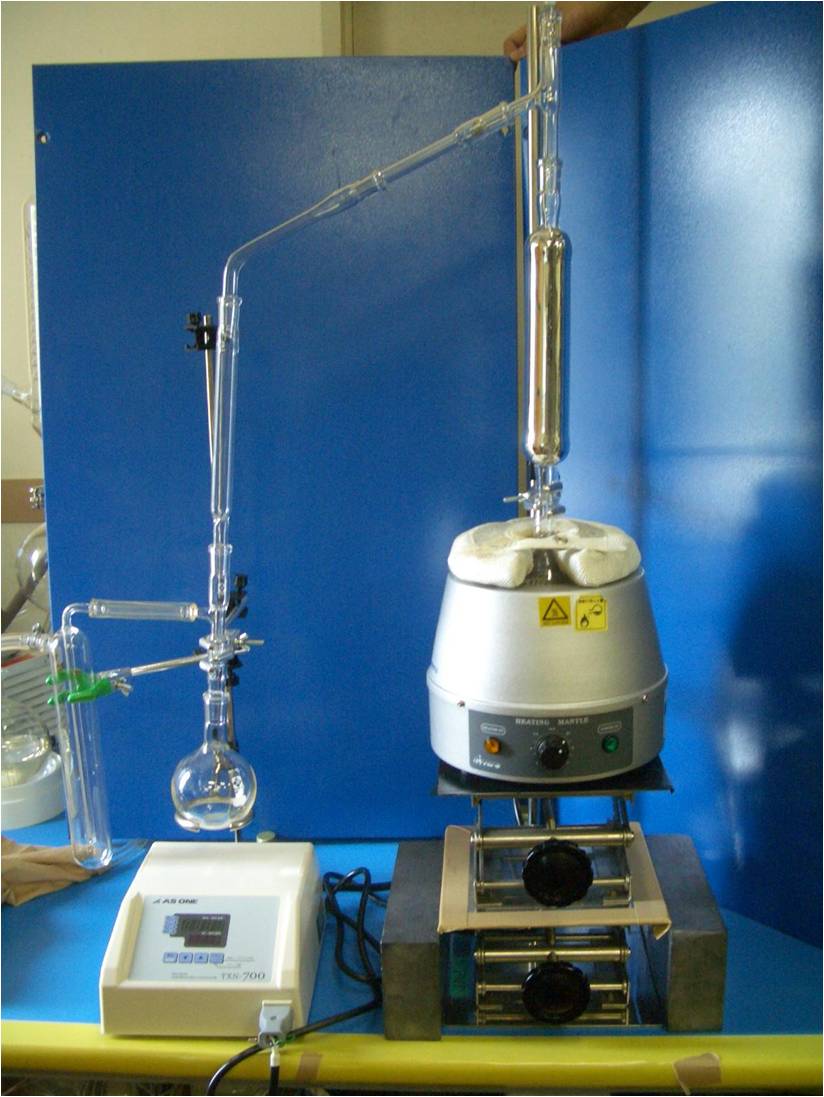}\label{Fig:TohDistillation}}
  \subfigure[]{\includegraphics[height=0.73\columnwidth]{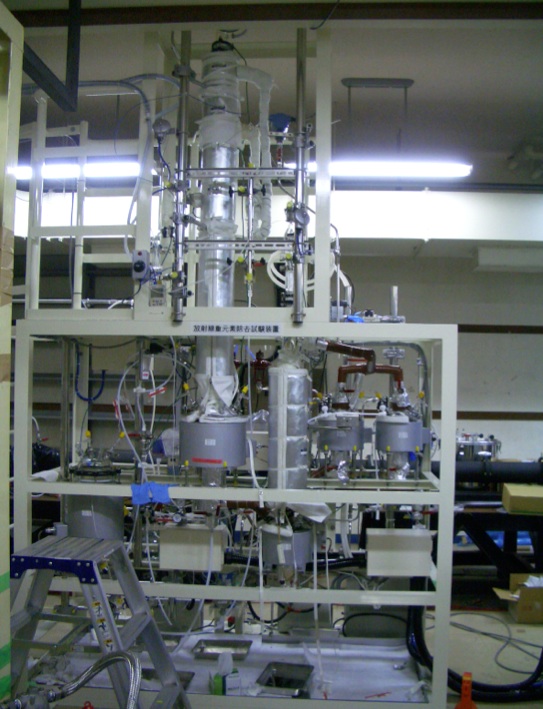}\label{Fig:TohLareDistllation}}
\caption{Size evolution of distillation systems utilized during the purification experiments. 
Fig.~\ref{Fig:UADistillation} and Fig.~\ref{Fig:TohDistillation} show the
bench-top systems used at the University of Alabama and Tohoku University,
respectively.  Together they provided a complete analysis of systematic studies
on distillation. Fig.~\ref{Fig:TohLareDistllation} is the scaled-up distillation
system at Tohoku University, allowing us to increase the amount of distillate by a factor 20.} 
\label{Fig:AllDistillation}
\end{center}
\end{figure*}

\input{Table07}

\subsection{Distillation}
\label{Sec:Distillation}

\par The final method studied for KamLAND's \ls\ purification addresses the removal of all \Lead\ species
in an integrated way: distillation.  During distillation, the distillate is heated to induce a phase 
change. Differences in vapor pressure result in fractionation between the constituents. 
As a by-product, the heating further breaks organo-metallic bonds and transforms organometallic \Lead\ compounds, 
that may be in solution, into a polar or ionic form. These, in turn, can be effectively removed by distillation.  

\par Distillation experiments primarily utilized \Lead-loaded $n$-dodecane, which makes up 80\% of the 
KamLAND \ls.  The distillation apparatuses utilized in these studies are
pictured in Fig.~\ref{Fig:AllDistillation}.   The system in 
Fig.~\ref{Fig:UADistillation} consisted of a three-neck flask (500 ml), a silvered 
fractionating column packed with stainless steel wool, and a condenser.  A thermocouple was used to regulate the distillate 
temperature to within a few~\degC. A mercury thermometer was placed on top of the fractionation column 
to monitor the temperature of the vapor prior to condensation.  The stainless steel wool 
increased the effective number of plates.  In essence, it
allows for multiple distillations of the liquid as the vapor rises 
through the column and re-condenses on the steel wool.  Typical operating
conditions were 170~hPa, with a $160 \pm 1$~\degC\ set-point for 
the distillate in the boiling flask.  The flask temperature was maintained by the thermocouple but the temperature measured by the mercury thermometer
 at the top of the fractionating column varied over time from 125-145~\degC.  Temperatures exceeding 145~\degC\ the top of the column 
resulted in a lower reduction factor due to the boil-over of contaminants.  The rate of distillation at these operating conditions was 10 ml/min. 
Operating the distillation system at a reduced pressure is important to minimize the amount of 
dissolved radioactive gases (\radon, \krypton, and \argon) as well as heat-induced changes in the optical properties of the distillate.

\par  Repeated distillation tests with \ThPb-loaded $n$-dodecane provided a handle on the variance associated with 
maintaining consistent operating procedures. To maintain reproducibility of the distillation results, the following procedures were utilized:

\begin{itemize}
  \item[-] All internal surfaces of the distillation system were cleaned with acetone, then
  etched with 0.1 - 1.0 molar HNO$_{3}$, and rinsed with DI water and ethanol.
  This eliminated carry-over between subsequent distillations.
  \item[-] All distillate collected before the top thermometer reached the boiling point of the distillate was discarded.  
    This removes all impurities with a low vapor pressure. Some \Lead\ candidates, addressed in this way, can be found in Table~\ref{Tab:OrganoPbProp}.
  \item[-] Collection of the distillate was stopped after approximately 90\% of the volume was distilled.  
    This operation reduces boil-over effects and re-contamination with
    high-boiling-point compounds.
  \item[-] A constant rate of distillation was maintained.  Rapid boiling at increased distillation rates directly
    contributes to the transfer of impurities into the distillate.
\end{itemize}

\par Distillation was performed under different pressure and temperature
conditions for $n$-dodecane, PC, and PPO.  Similar measurements taken with the
different distillation systems depicted in Fig.~\ref{Fig:AllDistillation}
allowed for the investigation of instrumental variations in the procedure that
could impact the reduction factor.
To ensure that the reduction factors observed in the modestly sized table-top experiments 
would apply to larger volumes, a scaled-up distillation system, shown in 
Fig.~\ref{Fig:TohLareDistllation}, was designed to process 5~L at a time.
Up to seven distillations were repeated in series to assess saturation of the \Lead\ reduction
factor.  Distillate samples were taken after consecutive
distillations during a single experiment.  The measured \ThPb\ reduction factors are 
summarized in Table~\ref{tab:DistRedFact}. Beyond three successive distillations, no further \Lead\ reduction could be observed. 
We also tested the combined effect of silica gel extraction and 5-times distillation
resulting in a factor 1.1 improvement compared to distillation only, as can be
seen in Table~\ref{tab:DistRedFact}. 

\par Distillation proved to be the most effective means of removing radioactive
\Lead\ from the KamLAND \ls\ components. The reduction factors observed during
the study of distillation ranged from a few hundred to several thousand,
depending on the operating conditions.  These results provided the necessary
technical base  required to achieve the high reduction factors needed for the
KamLAND purification campaign.  They further provided essential procedural
development and initial parameters for the full-scale KamLAND distillation system.

%% file: Table02.tex
\renewcommand\arraystretch{1.25}
\begin{table}[tdp]
    \caption{\ThPb\ reduction factors obtained with 250 ml of \ls\ utilizing
    different purification modes.  The uncertainties given for the Selecto
    results are statistical plus systematic added in quadrature.  The Aerosil
    results have only statistical uncertainties.  Tests with Selecto used 10 g of
    silica while only 0.5 g of Aerosil was used.}
  \begin{center}
    \begin{tabular*}{\columnwidth}{@{\extracolsep{\fill}} l l l}
      \hline
      \multicolumn{1}{c}{\textbf{Adsorbent}} & \multicolumn{2}{c}{\textbf{Reduction Factor}}\\
      \multicolumn{1}{c}{\textbf{}}   & \multicolumn{1}{c}{\textbf{Batch}} & \multicolumn{1}{c}{\textbf{Loop}} \\
      \hline \hline
      32-63 $\rm \mu$m Selecto   & 22.73 $ \pm 0.52 $  &  $ 4.98 \pm 0.12 $ \\
      100-200 $\rm \mu$m Selecto & 15.38 $ \pm 0.24 $  &  $ 7.04 \pm 0.10 $ \\
      Aerosil 200                &  3.45 $ \pm 0.02 $  &  $ 1.49 \pm 0.02 $ \\
      \hline
    \end{tabular*}
   \label{Ta:BatchLoop}
  \end{center}
\end{table}
\renewcommand\arraystretch{1.0}

%% file: Table03.tex
\renewcommand\arraystretch{1.25}
\begin{table*}[t!]
    \caption{Measured $^{212}$Pb reduction factors in \ls.  Only statistical
    uncertainties are quoted.  Experimental investigation of the systematic
    uncertainty for 32-64~$\mu$m gel yielded 7.2\%. The uncertainty on the mass
    measurements was 0.05~g.}
  \begin{center}
  \vskip 0.1in
  \begin{tabular*}{2.0\columnwidth}{@{\extracolsep{\fill}} p{3.0in} p{1.5in}@{} p{1.5in}@{}} 
      \hline
      \textbf {Adsorbent Type}  &  \textbf {\Lead\ Reduction Factor} & \textbf {Adsorbent Mass [g]}   \\ \hline \hline
      Selecto, Lot \#301286301, Si-gel 32-63 $\mu$m       & $ 19.4 \pm 0.47 $                      &  $ 5.0 $  \\  
      Selecto, Lot \#306279301, Si-gel 100-200 $\mu$m     & $ 15.38 \pm 0.24 $                     &  $ 10 $  \\  
      Selecto, Lot \#102085402, Alusil 70                 & $ 27.03 \pm 0.73 $                     &  $ 5.0 $  \\ 
      Selecto, Lot \#109110402, Alusil Plus               & $ 8.33 \pm 0.07 $                      &  $ 2.0 $  \\  
      Selecto, Lot \#108110403, Alusil NanoSmart          & $ 8.20 \pm 0.07 $                      &  $ 2.0 $  \\ 
      Selecto, Lot \#900110401, Si-gel NanoSmart ACT      & $ 3.58 \pm 0.01 $                      &  $ 2.0 $  \\ 
      Selecto, Lot \#900110401, Alusil Coarse             & $ 28.57 \pm 0.82 $                     &  $ 5.0 $  \\  
      Selecto, Lot \#107223405, Alusil 40 without K                      & $\rm 10.31 \pm 0.21 $      &  $ 2.0  $    \\ 
      \raggedright Aldrich, 3-(Diethylenetriamino) Propyl-Functionalized gel          & $\rm 8.33 \pm 0.69 $       &  $ 10.0 $   \\
      \raggedright Aldrich, Triamine Tetraacetate-Functionalized gel                  & $\rm 11.11 \pm 0.12 $      &  $ 10.2 $   \\
      Aerosil  200                                         & $\rm 8.26 \pm 0.07 $                      &  $ 0.5  $   \\
      S\"{u}d-Chemie,  Cu/Mn Catalyst T-2550              & $\rm 3.45 \pm 0.02 $                      &  $ 15  $    \\ 
      S\"{u}d-Chemie,  Cu/Mn Catalyst T-2550, Crushed     & $\rm 26.32 \pm 0.69 $                     &  $ 7.2 $ \\ 
      $\rm Ca_{3}(PO_{4})_{2}$                            & $\rm 6.54 \pm 6.54 $                      &  $ 2.0  $    \\ 
      \hline
    \end{tabular*}
  \end{center}
    \label{Ta:AdsorbentEff}
\end{table*}
\renewcommand\arraystretch{1.0}

%% file: Table04.tex
\renewcommand\arraystretch{1.25}
\begin{table*}[t!]
\caption[]{Chemical characteristics of the \ls\ components and of some known organometallic \Lead\ compounds. 
These data are taken from the CRC~\cite{bib:CRCChemPhys}.}
  \begin{center}
\begin{threeparttable}
  \begin{tabular*}{2.0\columnwidth}{@{\extracolsep{\fill}} l l c c }
    \hline
    \textbf {Name}  & \textbf {Formula}  & \textbf {Density [g/cm$^{3}$] @ 20~\degC} & \textbf {Boiling Point [~\degC]}  \\ 
    \hline \hline
    n-Dodecane                  &   $\rm C_{12}H_{26}$          &   0.7495     &    215            \\
    1,2,4-Trimethylbenzene (PC)  &   $\rm C_{9}H_{12}$           &   0.8758     &    169            \\
    2,5 Diphenyloxazole (PPO)   &   $\rm C_{15}H_{11}NO$        &   1.0940     &    360 \tnote{1}  \\
    Tetraethyllead              &   $\rm C_{8}H_{20}Pb$          &   1.653      &    200 \tnote{2}  \\
    Methyltriethyllead          &   $\rm C_{7}H_{18}Pb$         &   1.71       &    70             \\
    Diethyldimethyllead         &   $\rm C_{6}H_{16}Pb$         &   1.79       &    51             \\
    Ehtyltrimethyllead          &   $\rm C_{5}H_{14}Pb$         &   1.88       &    27             \\
    Tetramethyllead             &   $\rm C_{4}H_{12}Pb$         &   1.995      &    110            \\
    \hline  
  \end{tabular*}
\begin{tablenotes}
  \item[1] PPO has a melting point of 74~\degC  
  \item[2] Tetraethyllead will decompose at 200~\degC\ before it boils.
\end{tablenotes}
\end{threeparttable}
  \end{center}
\label{Tab:OrganoPbProp}
\end{table*}

%% file: Table05.tex
\renewcommand\arraystretch{1.5}
\begin{table}[t!]
  \caption[]{Combined \ThPb\ reduction factors for purification methods
  performed in series with an initial $\rm SiO_{2}$ purification. Secondary
  techniques specifically target \CPb\ bonds.  Only statistical uncertainties
  are reported except where noted.}
  \begin{center}
  \begin{threeparttable}
    \begin{tabular*}{\columnwidth}{@{\extracolsep{\fill}}  l c }
      \hline
      \multicolumn{1}{c}{\textbf{Method}}  & \textbf {Reduction Factor}      \\ \hline \hline
      $\rm SiO_{2}  $                                                &  $ 19.4 \pm 0.47 $  \tnote{1}           \\  
      $\rm SiO_{2} \rightarrow FeCl_{3} $                             &  $ > 1060  $                    \\  
      $\rm SiO_{2} \rightarrow FeCl_{3} \rightarrow SiO_{2} $         &  $ > 1263  $                     \\ 
      $\rm SiO_{2} \rightarrow SnCl_{4} $                             &  $ 47.0 \pm 1.1 $                \\  
      $\rm SiO_{2} \rightarrow MoS_{2}  $                             &  $ 24.9  \pm 0.6 $               \\   
      $\rm SiO_{2} \rightarrow Thiol Resin $                         &  $ 22.7 \pm 0.5 $                \\  
      $\rm SiO_{2} \rightarrow Distillation $                        &  $  > 4389  $  \tnote{2}          \\  
      \hline  
    \end{tabular*}
\begin{tablenotes}
  \item[1] Includes Systematic Uncertainties.
  \item[2] Experiment performed with n-Dodecane.
\end{tablenotes}
\end{threeparttable}
  \end{center}
  \label{tab:OrganicData}
\end{table}

\renewcommand\arraystretch{1.0}

%% file: Table06.tex
\renewcommand\arraystretch{1.5}
\begin{table}[t!]
\caption[]{ \ThPb\ reduction factors observed for heated \ls.  All heating was performed 
in a stainless steel container. The first column gives the heating time. The second column states the
average temperature during heating.  The final column provides the observed reduction factor.}
\begin{center}
\begin{threeparttable}
  \begin{tabular*}{\columnwidth}{@{\extracolsep{\fill}} c c c }
    \hline
    \textbf {Time [hrs]}  & \textbf {Avg. Temp [$\rm ^{o}C$]}  & \textbf {Reduction Factor}  \\ 
    \hline \hline
    \multicolumn{3}{c}{Heating + Silica Gel} \\
    2     & 142   &  $ 277.8 \pm 23.2 $  \tnote{1} \\
    2     & 187   &  $ 200  \pm 16 $  \tnote{1} \\ 
    1.5   & 160   &  $ 172.4 \pm 2.97 $  \tnote{2} \\ \hline
    \multicolumn{3}{c}{Heating Only} \\
    2     & 187   &  $ 5.3  \pm 0.06  $            \\
    2     & 157   &  $ 5.08  \pm 0.05  $            \\
    8     & 152   &  $ 9.35  \pm 0.09  $            \\ \hline
    \multicolumn{3}{c}{Control Sample, No Heat, Vessel Blank} \\
    3     & 17    &  $ 1.11  \pm 0.01  $ \\
    \hline  
  \end{tabular*}
\begin{tablenotes}
  \item[1] 5 g $\rm SiO_{2}$ in column.
  \item[2] 5 g of $\rm SiO_{2}$ and silica gel heated in column.
\end{tablenotes}
\end{threeparttable}
\end{center}
\label{tab:HeatingTable}
\end{table}
\renewcommand\arraystretch{1.0}

%% file: Table07.tex
\renewcommand\arraystretch{1.25}
 \begin{table*}[t!]
 \caption{ Average \ThPb\ reduction factors obtained by distillation of various
 components of \ls. The uncertainty is conservatively reported as the addition in quadrature of
 the error on the weighted mean with the standard deviation of the measurements
 to account for the large system-to-system variation relative to the individual
 measurement uncertainties.}
  \begin{center}
     \begin{tabular*}{1.4\columnwidth}{@{\extracolsep{\fill}} l c }
        \multicolumn{1}{c}{\textbf{Distilled material}} & \multicolumn{1}{c}{\textbf{Averaged Reduction Factors}} \\
        \hline\hline
        \multicolumn{1}{c}{\textbf{$n$-Dodecane}} & \multicolumn{1}{c}{\textbf{300 hPa, 160~\degC, Bench-top System}~\cite{bib:TakeuchiMastersThesis, bib:MomoyoMastersThesis}}  \\
        \hline\hline
        1 time distillation  & $(3.4 \pm 2.3)\times10^3$  \\ 
        3 times distillation & $(1.9 \pm 1.0)\times10^4$  \\ 
        5 times distillation & $(2.2 \pm 0.6)\times10^4$  \\ 
        7 times distillation & $(2.8 \pm 0.6)\times10^4$  \\ 
        \hline
        \multicolumn{1}{c}{\textbf{$n$-Dodecane}} & \multicolumn{1}{c}{\textbf{190 hPa, 160~\degC, Bench-top System}}  \\
        \hline\hline
        1 time distillation                    & $(7.8 \pm 3.2)\times10^2$             \\    
        5 times distillation                   & $(2.6 \pm 0.2)\times10^3$  \\    
        5 times distillation + SiO$_2$         & $(2.9 \pm 0.6)\times10^3$  \\    
        \hline
        \multicolumn{1}{c}{\textbf{$n$-Dodecane}} & \multicolumn{1}{c}{\textbf{37 hPa, 113~\degC, Scaled-up System}~\cite{bib:TakeuchiMastersThesis}}  \\
        \hline\hline
        1 time distillation   & $(1.4 \pm 0.3)\times10^3$         \\
        3 times distillation  & $(3.8 \pm 1.2)\times10^3$        \\
        5 times distillation  & $> 1.3\times10^4$    \\
        \hline
        \multicolumn{1}{c}{\textbf{Pseudocumene}} & \multicolumn{1}{c}{\textbf{37 hPa, 75~\degC, Scaled-up System}~\cite{bib:KibeMastersThesis}}  \\
        \hline\hline
        1 time distillation  & $(7.7 \pm 3.2)\times10^2$\\
        2 times distillation & $(1.3 \pm 1.2)\times10^3$\\
        3 times distillation & $(2.8 \pm 1.3)\times10^3$\\
        4 times distillation & $(7.8 \pm 4.8)\times10^3$\\
        \hline
        \multicolumn{1}{c}{\textbf{PPO}} & \multicolumn{1}{c}{\textbf{20 hPa, 160~\degC, Bench-top System}~\cite{bib:MomoyoMastersThesis}}  \\
        \hline\hline
        1 time distillation                    & $(2.1 \pm 0.8)\times10^3$   \\
        \hline
    \end{tabular*}
   \end{center}
\label{tab:DistRedFact}
\end{table*}
 
\renewcommand\arraystretch{1.0}

%% file: Conclusion.tex
\section{Conclusion}
\label{Sec:Conclusion}

\par  The removal of \Radon-born \Lead\ contamination from organic liquid scintillator was studied in preparation
of a low background phase of the KamLAND experiment. Reduction factors were measured for various purification
methods: water extraction with purified water and ion-enhanced water, isotope
exchange, filtration, adsorption, and distillation.

\par Among these methods, distillation was identified as the
most efficient method. It provides the reduction factors
of 10$^3$ -- 10$^4$ required for KamLAND's solar neutrino detection phase.
We obtained reduction factors for all components of the
KamLAND LS: $n$-dodecane, pseudocumene and PPO.
The studies performed on a scaled-up system provide
the basic data for the design and operation of a purification
plant for continuous circulation and purification of 1000~tons of KamLAND LS.

\par The second most effective procedure was silica gel
extraction, which provided highly reproducible results.
However, a steady decline observed in lead reduction
factors after successive extractions cannot be explained
by simple plate theory. One possible explanation
is that a portion of the dissolved lead exists
in non-polar organometallic compounds, which can be
degraded by heat and/or chemically targeted on the \CPb\ 
bond. Our studies on silica gel extraction with preheating
and processing with FeCl$_3$ showed improved reduction
factors consistent with the existence of such non-polar
lead compounds in LS.